\documentclass[
 aip,
 amsmath,amssymb,
 reprint,%
]{revtex4-1}
\usepackage[colorlinks=true,citecolor=blue]{hyperref}
\usepackage{graphicx}
\usepackage{dcolumn}
\usepackage{bm}
\usepackage[normalem]{ulem}
\usepackage{calrsfs}
\DeclareMathAlphabet{\pazocal}{OMS}{zplm}{m}{n}
\usepackage{appendix}

\usepackage{soul,xcolor}
\setstcolor{red}

\newcommand{\be}{\begin{equation}}
\newcommand{\ee}{\end{equation}}
\newcommand{\ba}{\begin{eqnarray}}
\newcommand{\ea}{\end{eqnarray}}
\newcommand{\tg}{\tilde g_{\mu\nu}}

\newcommand{\h}{\textsl{H}}

\usepackage[utf8]{inputenc}
\usepackage[T1]{fontenc}
\usepackage{mathptmx}
\usepackage{etoolbox}

\makeatletter
\def\@email#1#2{%
 \endgroup
 \patchcmd{\titleblock@produce}
  {\frontmatter@RRAPformat}
  {\frontmatter@RRAPformat{\produce@RRAP{*#1\href{mailto:#2}{#2}}}\frontmatter@RRAPformat}
  {}{}
}%
\makeatother

\begin{document}

\title[GWs and BH perturbations in Acoustic Analogues]{Gravitational Waves and Black Hole perturbations in Acoustic Analogues}

\author{Chiara Coviello}
\email{chiara.coviello@kcl.ac.uk}
\affiliation{Department of Physics, King’s College London, The Strand, London, WC2R 2LS, UK}

\author{Maria Luisa Chiofalo}
\affiliation{Dipartimento di Fisica, Universit\`a di Pisa, Polo Fibonacci, Largo B. Pontecorvo 3, 56127  Pisa, Italy}
\address{INFN, Sezione di Pisa,
Polo Fibonacci, Largo B. Pontecorvo 3, 56127 Pisa, Italy}

\author{Dario Grasso}
\affiliation{INFN, Sezione di Pisa,
Polo Fibonacci, Largo B. Pontecorvo 3, 56127 Pisa, Italy}

\author{Stefano Liberati}
 \affiliation{Scuola Internazionale Superiore Studi Avanzati (SISSA), Physics Area, Via Bonomea 265, 34136 Trieste, Italy
}
\affiliation{Institute for Fundamental Physics of the Universe (IFPU), Via Beirut 2, 34014 Trieste, Italy}
\affiliation{INFN,  Sezione di Trieste, Via Valerio 2, 34127 Trieste, Italy}

\author{Massimo Mannarelli}
\affiliation{INFN, Laboratori Nazionali del Gran Sasso, Via G. Acitelli,
22, I-67100 Assergi (AQ), Italy}

\author{Silvia Trabucco}
\affiliation{Gran Sasso Science Institute, Viale Francesco Crispi, 7, 67100 L'Aquila, Italy}
\affiliation{INFN, Laboratori Nazionali del Gran Sasso, Via G. Acitelli,
22, I-67100 Assergi (AQ), Italy}

\date{21 February 2025}

\begin{abstract}
Phonons in Bose-Einstein condensates propagate as massless scalar particles on top of an emergent acoustic metric.  This hydrodynamics/gravity analogy can be exploited to realize  acoustic black holes, featuring an event horizon that traps phonons. We show that by an appropriate external potential, gravitational wave-like perturbations of the acoustic metric  can be produced. Such perturbations can be used to excite an acoustic black hole, which should then relax by phonon emission.
\end{abstract}

\maketitle

\section{\label{sec:intro}Introduction}
Besides their interest as prominent astrophysical objects, black holes (BHs) serve as perfect playgrounds for theoretical physicists interested to explore the deepest and possibly intertwined aspects of quantum mechanics and general relativity. Indeed, a full understanding of BHs is likely to require insights from both fields. This line of research  is of great physical interest, because  solving   open problems regarding the nature of BHs  could lead us to significantly improve our comprehension of gravity.\\
The detection of gravitational waves (GWs) originating from coalescing binary BHs  has opened up new avenues for probing these objects \cite{2016:GWs}. This discovery holds also the potential to reveal, through future generation detectors, aspects of their quantum mechanical nature \cite{2024:LQGandGW}. However, like other astronomical observations, it is inherently constrained by the challenge of studying uncontrolled physical systems in the cosmos.\\
Recently, rapid advancements in quantum technologies have begun to address this limitation by enabling the creation of analogue (sonic) BHs in laboratory settings, offering a controlled environment to explore similar phenomena. These systems, typically created through quantum fluids, share the same kinematic properties as their astrophysical counterparts by mimicking the same metric, while the dynamics of the background will generally differ. The connection between hydrodynamics and gravity lies in the fact that the low-energy phonons propagate as  massless scalar fields on an effective acoustic metric. Once the metric is determined, the comparison between sonic and astrophysical systems can provide valuable insights into specific gravitational theories\cite{Barcelò_analoguegravityReview}. One of the most remarkable achievements of this approach has been the quantum simulation of the Hawking radiation  from a sonic horizon in a Bose-Einstein condensate (BEC) \cite{2016Steinhauer} implementing the idea proposed by Unruh for a conventional fluid \cite{Unruh_experimentalBHev}.
Although sonic BHs are clearly very different from astrophysical BHs -- most importantly the former do not obey Einstein equations -- they can however test {\it universal} properties common to both entities.
For example in Ref.\,\onlinecite{CGMT_1Kinetic} it has been shown that, similarly to what computed for astrophysical BHs, the expression of the Hawking temperature of a sonic BH in a BEC springs from quantum mechanics and the assumption of the area law for the horizon entropy. \\
Following that line of reasoning, we employ analogue systems to explore open questions about sonic BHs and gravity. In particular, we show how specifically designed analogue systems can be used  to mimic the propagation of GWs in a BEC. In general relativity, GWs can be treated as classical perturbations of the background metric. Here, we mimic such metric perturbations employing appropriate external potentials. 
We will refer to the variations of the acoustic metric produced by such external potentials as GW-like perturbations. We discuss the propagation of GW-like perturbations onto two different metrics: vacuum and analogue BH.
The study of GW-like perturbations in hydrodynamic systems was first discussed in Ref.\,\onlinecite{Hartley_GWsimulation}, which focused on the interaction, and possible detection, of real GWs with BECs. 
Here we  treat complex and interesting physical situations as that of a GW-like perturbation impinging onto a sonic BH. The systems discussed in this paper pave the way for investigating significant theoretical questions that can also be experimentally tested.\\
This paper is organised as follows. In Sec.\,\ref{sec:1}, we review the acoustic metric in the context of non-relativistic BECs. Sec.\,\ref{sec:GW_flat} demonstrates how to simulate a GW perturbation within a flat background acoustic metric in a BEC. We then extend this analysis in Sec.\,\ref{sec:3}, where we examine how to create  GW-like perturbations on the acoustic BH metric within a BEC environment. At the beginning of this section, we introduce a vortex geometry, which serves as the foundation for our model of an acoustic BH disturbed by GW-like perturbations. Finally, in Sec.\,\ref{sec:conclusions}, we summarize our findings and discuss potential future research directions enabled by our models.

\section{\label{sec:1}Theoretical setup}
In this section we briefly discuss the emergence of the acoustic metric in non-relativistic BECs  at vanishing temperature\cite{Barcelò_analoguegravityReview}. This is achieved by an appropriate scale separation between the background fluid and the low-energy (phonon) excitations.\\
For a Bose gas in the dilute gas approximation, where the average spacing between particles is greater than the scattering length $a$, the true interaction potential between particles can be approximated using the Fermi pseudo-potential. Under this approximation and employing the Hartree-Fock method, the total wavefunction of the system, consisting of $N$ bosons, can be represented as a product of single-particle functions.  This allows us to describe the diluted Bose gas through a single quantum field, $\hat{\Psi}(t,\mathbf{x})$, which satisfies the Gross-Pitaevskii equation \cite{Dalfovo:1999zz}:
\begin{equation}
    i\hslash \frac{\partial}{\partial t}\hat{\Psi}=\biggl(-\frac{\hslash^2}{2m}\nabla^2 + V_{\text{ext}}(\mathbf{x})+\gamma(a) \hat{\Psi}^{\dag}\hat{\Psi}\biggr) \hat{\Psi},
\label{eq:GrossQunatumFiledEntire}
\end{equation}
where $m$ is the mass of the bosons, $V_{\text{ext}}$ is the external potential and \be \gamma(a)=\frac{4\pi a\hslash^2}{m}\,, \ee parameterises the strength of the interaction between bosons.
The quantum field can be divided into a macroscopic (classical) condensate $\psi(t,\mathbf{x})=\langle\hat{\Psi}(t,\mathbf{x})\rangle$, and a fluctuation $\hat{\varphi}(t,\mathbf{x})$: 
\begin{equation}
    \hat{\Psi}(t,\mathbf{x})=\psi(t,\mathbf{x})+\hat{\varphi}(t,\mathbf{x}),
\end{equation}
where the expectation value is defined as the trace of the density matrix times the operator itself. This splitting indicates a separation of scales, with the fluctuation $\hat{\varphi}(t,\mathbf{x})$ being linked to rapid and short-distance  quantum fluctuations of the wavefunction, while the background field $\psi(t,\mathbf{x})$ accounts for the large-distance and long-time scale variations. Neglecting back-reactions of the quantum fluctuations on the background, the equation for the classical wave function is given by
\begin{equation}
    i\hslash \frac{\partial}{\partial t}\psi=\biggl(-\frac{\hslash^2}{2m}\nabla^2 + V_{\text{ext}}(\mathbf{x})+\gamma(a) n_c \biggr) \psi,
\label{eq:gross1}
\end{equation}
with $n_c\equiv |\psi(t,\mathbf{x})|^2$ the condensate density. Equation\,\eqref{eq:gross1} is the most commonly adopted expression to approximate the dynamics of the condensate wavefunction. The linearized quantum fluctuations, on the other hand, follow the Bogoliubov-de Gennes equation: 
\begin{align}\label{eq:BdG}
    i\hslash \frac{\partial}{\partial t}\hat{\varphi}(t,\mathbf{x})=&\biggl(-\frac{\hslash^2}{2m}\nabla^2 + V_{\text{ext}}(\mathbf{x})+\gamma(a)2 n_c \biggr)\hat{\varphi}(t,\mathbf{x})\nonumber\\ &+\gamma(a) \psi^2(t,\mathbf{x}) \hat{\varphi}^{\dag}(t,\mathbf{x}),
\end{align}
which depends on  the solution of mean-field equation. This set of equations arises from neglecting back-reactions within the self-consistent mean-field approximation. Would we account for the back-reaction of quantum fluctuations on the dynamics of the condensate wavefunction, the equations governing the mean-field and quantum fluctuations become inseparable.\\
In order to describe the quantum fluctuations as massless scalar field propagating on top of an emergent acoustic metric, we use the Madelung representation for the condensate wavefunction
\begin{equation}
    \psi(t,\mathbf{x})=\sqrt{n_c(t,\mathbf{x})}e^{-i\theta(t,\mathbf{x})/\hslash},
\end{equation}
and the quantum acoustic representation for the fluctuation
\begin{equation}
    \hat{\varphi}(t,\mathbf{x})=e^{-i\theta/\hslash}\left(\frac{1}{2\sqrt{n_c}}\hat{n}_1-i\frac{\sqrt{n_c}}{\hslash}\hat{\theta}_1\right)\,,
\end{equation}
where $\hat{n}_1$ and $\hat{\theta}_1$ are two real scalar fields corresponding to phonon
density and phase fluctuations, respectively \cite{Barcelò_analoguegravityReview}. Upon substituting this expression in Eq.\,\eqref{eq:BdG} we obtain two coupled differential equations for $\hat{n}_1$ and $\hat{\theta}_1$. These equations can be cast as a single nonlocal differential equation for $\hat{\theta}_1$, see Ref.\onlinecite{Barcelò_analoguegravityReview} for more details.  In the hydrodynamic limit, that is for length scales much larger than the healing length\,\cite{Dalfovo:1999zz}, such equation can be written as that of a quantum scalar field propagating over a curved background: 
\begin{equation}
    \square \hat{\theta}_1=\frac{1}{\sqrt{|g|}}\partial_{\mu}(\sqrt{|g|}g^{\mu\nu}\partial_{\nu}\hat{\theta}_1)=0,
\end{equation}
with the so-called acoustic metric
\begin{equation}
\label{eq:analogue_metric_BEC}
    g_{\mu\nu}(t, \mathbf{x}) = \frac{n_c}{mc_s}
    \begin{pmatrix}
        -[c_s^2 - v^2] & -v_j \\
        -v_i & \delta_{ij}
    \end{pmatrix}
    ,
\end{equation}
where \be\label{eq:c_s} c_s=\sqrt{\frac{\gamma(a) n_c}{m}}\,,\ee is the adiabatic speed of sound and
\be\label{eq:potential_velocity} \mathbf{v}=\frac{\nabla \theta}{m}\,,
\ee
is the irrotational velocity field of the condensate. The $00$ component of the metric vanishes when the fluid velocity equals the speed of sound; this corresponds to the acoustic horizon.

\section{\label{sec:GW_flat}Gravitational Wave acoustic metric}
So far we have seen that the propagation of the phonon field is formally equivalent to the propagation of a massless scalar field in curved spacetime, with the acoustic metric given in Eq.\,\eqref{eq:analogue_metric_BEC}. We now consider to which extent the perturbations of the acoustic metric can be described as GWs. Notice that we consider two different types of perturbations: background perturbations, which are externally driven and correspond to the acoustic metric variations, and phonons, which are the quantum fluctuations of the system and can be described as massless modes propagating on the emergent acoustic metric. The  separation between these two types of perturbations is necessary and can be ensured because the frequency of the external perturbation, being controllable, can be chosen arbitrarily. \\
We begin with considering GWs propagating in flat spacetime  (GWs in a BH-like metric are instead discussed in the next section).  We  first briefly recap how GWs are introduced in general relativity, then we show that GWs expressed in a particular gauge can be emulated by  appropriate fluid perturbations. 

\subsection{Gravitational waves in flat spacetime}
The metric we aim to reproduce is that of GWs in vacuum\cite{Maggiore:2007ulw}:
\begin{equation}
\label{eq:linexpGW_1}
    \gamma_{\mu\nu}=\eta_{\mu\nu}+\epsilon\, h_{\mu\nu},
\end{equation}
where $\epsilon \ll 1$ and we indicate with $\gamma_{\mu\nu}$ the gravitational metric tensor, $\eta_{\mu\nu}$  the Minkowski metric, while $h_{\mu\nu}$ represents the metric perturbation. GWs are metric perturbations solution of the Einstein's field equations in the linearized theory. 
An important symmetry in this linearized framework is represented by coordinate (or gauge) transformations of the form:
\begin{equation}
    \label{eq:gaugetrasf_lin}
    x^\mu \to x^{\prime\mu}=x^\mu+\epsilon'\zeta^\mu(x),
\end{equation}
where $\epsilon' \ll 1$ and $\zeta^\mu(x)$ a generic four-vector. This gauge transformation induces a metric transformation, which can be cast as a metric perturbation. Taking $\epsilon=  \epsilon'$, at the first order in $\epsilon$, this transformation is:
\begin{align}
        \label{eq:gaugetrasf_hmunu}
    \eta_{\mu\nu}+\epsilon h_{\mu\nu}(x)\to&\eta_{\mu\nu}+\epsilon h^{\prime}_{\mu\nu}(x^\prime)\nonumber\\&=\eta_{\mu\nu}+\epsilon \left( h_{\mu\nu}-\partial_\mu\zeta_\nu-\partial_\nu \zeta_\mu \right).
\end{align}
Thus, slowly varying infinitesimal diffeomorphisms are  symmetries of linearized theory of gravity. In contrast, full general relativity has complete invariance under all coordinate transformations, not limited to infinitesimal changes.\\
By an appropriate gauge transformation it is possible to set   $h^\mu{}_\mu=0$, then in the Lorentz gauge 
\be
\partial^\nu h_{\mu\nu}= 0\,,
\ee
the in vacuum  propagation of the GW is described by the wave equation
\begin{equation}
\label{eq:Einstein_lin_nosource}
(-\partial_t^2+c^2\nabla^2)h_{\mu\nu}=0\,,
\end{equation}
indicating that GWs propagate at the speed of light $c$.\\
One may further simplify the treatment of GWs using the transverse-traceless (TT) gauge, in this case the metric perturbation  satisfies the conditions:
\begin{equation}
    h^{0\mu}=0, \quad h^i{}_i=0,\quad \partial^j h_{ij}=0\,,
\end{equation}
hence in the TT gauge a GW propagating along the $z$ axis is  described by
\begin{align}
    \label{eq:gw_tt}
    h_{\mu\nu}^{TT}(t,z)=&
    \begin{pmatrix}
     0 &0&0&0\\
     0& h^{TT}_{+}(t,z) & h^{TT}_{\times}(t,z) &0 \\
     0& h^{TT}_{\times}(t,z)& -h^{TT}_{+}(t,z) &0 \\
     0 &0&0&0\\
    \end{pmatrix}\nonumber\\ =&
    \begin{pmatrix}
     0 &0&0&0\\
     0& h_{+} & h_{\times} &0 \\
     0& h_{\times}& -h_{+} &0 \\
     0 &0&0&0\\
    \end{pmatrix}
    \cos \left(\omega(t-z/c)\right),
\end{align}
where $h_{+}$ and $h_{\times}$ denote the amplitudes of the ``plus'' and ``cross'' polarizations of the wave. 
As we will see, it is not possible to express the acoustic metric perturbation in this form. However, given the simplicity of the GWs in TT-gauge, we will start from this gauge to obtain an expression of $h_{\mu\nu}$ that can be matched to the acoustic metric perturbation.

\subsection{Method}
In order to emulate a GW propagating in flat spacetime with a BEC, we have to cast the acoustic metric in Eq.\,\eqref{eq:analogue_metric_BEC} as in Eq.\,\eqref{eq:linexpGW_1}, that is a Minkowski background with a propagating GW perturbation. To this end, we have to determine how the BEC background perturbations change the acoustic metric and then look for a gauge transformation of the space coordinates such that $h^{\mu\nu}$ takes a form that can be matched to the acoustic metric perturbation.\\
More in details, the emulation of the GW with an acoustic analogue is obtained with the following steps:
\begin{enumerate}
    \item \textit{Acoustic metric perturbation}:  Perturb the fluid background quantities to induce an acoustic metric perturbation. For small perturbations  the metric  in Eq.\, \eqref{eq:analogue_metric_BEC} can then be written  as a background $g_{\mu\nu}^0$ plus a perturbation $\tilde g_{\mu\nu}$: \be
    g_{\mu\nu}=g_{\mu\nu}^0+\epsilon\tilde g_{\mu\nu}\,.\ee
    \item \textit{GWs in different gauges}: Exploit the gauge symmetry of general relativity, to find the   metric of a GW that propagates in vacuum $h_{\mu\nu}$ in a form that can be matched to $\tilde g_{\mu\nu}$.
    \item \textit{Comparison of  GW with the perturbed acoustic metric}: Identify the fluid characteristics needed to match  $g_{\mu\nu}^0$  with $\eta_{\mu\nu}$ and $\tilde g_{\mu\nu}$ with $h_{\mu\nu}$.
    \item \textit{Physical systems:} Check whether 
    the choice of the condensate quantities needed to match the acoustic metric with the gravitational metric are physical, meaning that the continuity and the Euler equations, as well as the irrotational condition are all satisfied. To this aim, we introduce an external potential which mimics a GW.
\end{enumerate}
This approach is inspired by the work in Ref.\onlinecite{Hartley_GWsimulation}, where a propagating homogeneous GW is simulated in a BEC. The main difference is that they do not consider the spatial dependence of the GW, while we do. This introduces technical complications, since  the space dependence does not make it possible to work in Fermi normal coordinates\cite{Maggiore:2007ulw} for the step 2, i.e. the inertial limit of the proper detector frame, as done in Ref.\, \onlinecite{Hartley_GWsimulation}. Because of that, we need to introduce a different suitable gauge, which is not the usual coordinate reference frame.

\subsubsection{Acoustic metric perturbation}
\label{sec:perturbedacousticmetric-sec}
The acoustic metric in Eq.\, \eqref{eq:analogue_metric_BEC} 
depends on the number density, the fluid velocity and  speed of sound. Now we assume that there is an external potential that determines a space-time variation of these quantities:
\begin{equation}
    \begin{aligned}
        &n_c \to n_c+\epsilon\,\delta n_c \\
        &v_{i} \to v_{i}+\epsilon\,\delta v_{i}\\
        &c_s \to c_s +\epsilon\, \delta c_s\,,
    \end{aligned}
\label{eq:backg_fluctuations}
\end{equation}
on top of their corresponding averages $n_c$, $v_i$ and $c_s$. Upon plugging the varied quantities in Eq.\, \eqref{eq:analogue_metric_BEC} one obtains a slight change of the acoustic metric.
Using  the coordinates $x^{\mu}=(c_s t,x,y,z)$, the unperturbed metric is given by
\begin{equation}
    \label{eq:analog_eta_dx0=csdt}
    g_{\mu\nu}^0= \frac{n_c}{m c_s}
    \begin{pmatrix}
        -\left(1-\frac{v^2}{c_s^2}\right) &  -\frac{v_{j}}{c_s}\\
        -\frac{v_{i}}{c_s} & \delta_{ij}\\
    \end{pmatrix}\,,
\end{equation}
while the metric perturbation at first order in $\epsilon$ is
\begin{equation}
    \label{eq:analog_h_dx0=csdt}
    \tilde g_{\mu\nu}(t,\mathbf{x})= \frac{n_c}{m c_s}
\begin{pmatrix}
    \tilde g_{00} & \tilde g_{01} &\tilde g_{02} & \tilde g_{03}\\
    \tilde g_{01} &\tilde g_{11}&0&0\\
    \tilde g_{02} &0&\tilde g_{22}&0\\
    \tilde g_{03} &0&0&\tilde g_{33}\\
\end{pmatrix}
\,,
\end{equation}
where the metric components are:
\begin{align}
    \tilde g_{00}&=2\frac{v^2}{c_s^2}\frac{\delta v}{v}-\frac{1}{2}\frac{(c_s^2+v^2)}{c_s^2}\frac{\delta a}{a}-\frac{1}{2}\frac{3c_s^2-v^2}{c_s^2}\frac{\delta n_c}{n_c}\,,\\
    \tilde g_{0i}&=-\frac{1}{2}\frac{v_{i}}{c_s}\frac{\delta n_c}{n_c}+\frac{1}{2}\frac{v_{i}}{c_s}\frac{\delta a }{a}-\frac{v_{i}}{c_s}\frac{\delta v_{i}}{v_{i}}\,,\\
    \tilde g_{ij}&=\frac{1}{2}\left(\frac{\delta n_c}{n_c}-\frac{\delta a}{a}\right)\delta_{ij}.
\label{eq:tgig}
\end{align}
with $\delta c_s/c_s=\delta a/(2 a) + \delta n_c/(2 n_c)$. Apart from a prefactor, $g_{\mu\nu}^0$ written in Eq.\,\eqref{eq:analog_eta_dx0=csdt} is the flat spacetime metric only if $v_i=0$ for all $i$. With this choice, the perturbation of the acoustic metric becomes:
\begin{equation}
    \label{eq:acousticPerturbMetric}
    \tilde g_{\mu\nu}=\frac{n_c}{m c_s}
    \begin{pmatrix}
        -\frac{3}{2}\frac{\delta n_c}{n_c}-\frac{1}{2}\frac{\delta a}{a}& -\frac{\delta v_j}{c_s}\\
        -\frac{\delta v_i}{c_s}& \left(\frac{1}{2}\frac{\delta n_c}{n_c}-\frac{1}{2}\frac{\delta a}{a}\right)\delta_{ij}\\
    \end{pmatrix}
    \,.
\end{equation}
At this point, we have to find a gauge in which the GW metric can be written in the same form of the metric in Eq.\,\eqref{eq:acousticPerturbMetric}.

\subsubsection{\label{sec:GW_newgauges-sec}GWs in different gauges}
In this section, our goal is to derive an expression for $h_{\mu\nu}$, representing a GW metric, that can be matched to $\tg$. We use the symmetry of linearized general relativity, as described in Eq.\,\eqref{eq:gaugetrasf_lin}, which induces the transformation of $h_{\mu\nu}$ outlined in Eq.\eqref{eq:gaugetrasf_hmunu}. The traditional TT gauge used for GWs (refer to Eq.\,\eqref{eq:gw_tt}) does not align with $\tg$ see Eq.\,\eqref{eq:acousticPerturbMetric}, nor do Fermi normal coordinates\cite{Maggiore:2007ulw}. Consequently, we have to adopt a different reference frame, distinct from both the TT gauge and the local detector frame. We start from the TT gauge and then perform a linearized gauge transformation of the type given  in Eq.\,\eqref{eq:gaugetrasf_lin} with
\begin{align}
    \zeta_0&=(lx+by+dz)(fh_+ + g h_{\times}) \cos \left(\omega(t-z/c)\right)\nonumber\,,\\
    \zeta_x &=\frac{1}{2}(x h_{+}+y h_{\times})\cos \left(\omega(t-z/c)\right)\nonumber\,,\\
    \zeta_y&=\frac{1}{2}(x h_{\times}-y h_+)\cos \left(\omega(t-z/c)\right)\nonumber\,, \\
    \zeta_z&=0\,,
\label{eq:CC_gauge_1}
\end{align}
where $h_{+}$ and $h_{\times}$ are the amplitudes of the GW in the TT gauge and  $l,b,d,f,g$ are real numbers. In this gauge the linearized Einstein's equations do not in general take the simple form given in Eq.\,\eqref{eq:Einstein_lin_nosource}. However, we do not need to determine such equations, because we know that under a gauge transformation  $h_{\mu\nu}$ transforms according to Eq.\,\eqref{eq:gaugetrasf_hmunu}. Hence, plugging Eqs. \,\eqref{eq:CC_gauge_1}  in Eq.\,\eqref{eq:gaugetrasf_hmunu}
we derive the metric perturbation at first order in $\epsilon$, $h'_{\mu\nu}(x^{\prime\rho})$. Its  nonvanishing components are
\begin{equation}
    h'_{13}(x^{\prime \rho})=-\frac{\omega}{2c}(x' h_+ + y' h_{\times})\sin(\omega(t'-z'/c))\,,
\end{equation}
\begin{equation}
    h'_{23}(x^{\prime \rho})=-\frac{\omega}{2c}(x' h_{\times}-y' h_{+})\sin(\omega(t'-z'/c))\, ,
\end{equation}
\begin{align}
    h'_{00}(x^{\prime \rho})=&
    \frac{2\omega}{c}(lx'+by'+dz')(fh_+ + g h_{\times})\nonumber\\ &\times\sin(\omega(t'-z'/c))\,,
\end{align}
\begin{align}
   h'_{01}(x^{\prime \rho})=&\frac{\omega}{2c}(x'h_+ + y'h_{\times})\sin(\omega(t'-z'/c))\nonumber\\&-l(fh_+ + gh_{\times})\cos(\omega(t'-z'/c))\,,
\end{align}
\begin{align}
    h'_{02}(x^{\prime \rho})=&\frac{\omega}{2c}(x'h_{\times} - y'h_{+})\sin(\omega(t'-z'/c))\nonumber\\&-b(fh_+ + gh_{\times})\cos(\omega(t'-z'/c))\,,
\end{align}
\begin{align}
    h'_{03}(x^{\prime \rho})=&-d(fh_+ + gh_{\times})\cos(\omega(t'-z'/c))-\frac{\omega}{c}\nonumber\\&\times(lx'+by'+dz')(fh_+ + gh_{\times})\sin(\omega(t'-z'/c))),
\end{align}
with $x^{\prime\rho}=(ct',x',y',z')$.  
Therefore, in this gauge the GW is given by
\be\label{eq:hmunu_matrix}
h'_{\mu\nu}= \begin{pmatrix}
    h'_{00} & h'_{01} &h'_{02} &  h'_{03}\\
    h'_{01} &0 &0&h'_{13}\\
    h'_{02} &0&0&h'_{23}\\
   h'_{03} &h'_{13}&h'_{23}&0
\end{pmatrix}\,,
\ee
which should be compared with $\tg$ in Eq.\,\eqref{eq:analog_h_dx0=csdt}. 
The difference is that in this gauge  $h'_{13},h'_{23}\not=0$, which should 
instead vanish, while $h'_{11} = h'_{22} =h'_{33}=0$, while they should not vanish.
We note that if we assume \be \frac{x'\omega}c\ll 1  \quad \text{and} \quad \frac{ y'\omega}c \ll 1\,,\label{eq:long}\ee   which we henceforth refer to as the long wavelength limit, $h'_{13}$ and $h'_{23}$ are suppressed and the the dominant terms are: 
\begin{align}
    \label{eq:trasf1_h00}
    h'_{00}(x^{\prime \rho})=&2dz'\frac{\omega}{c}(fh_++g h_\times)\sin(\omega(t'-z'/c))\,,\\
    \label{eq:trasf1_h01}
    h'_{01}(x^{\prime \rho})=&-l(fh_+ + gh_{\times})\cos(\omega(t'-z'/c))\,,\\ 
    \label{eq:trasf1_h02}
    h'_{02}(x^{\prime \rho})=&-b(fh_+ + gh_{\times})\cos(\omega(t'-z'/c))\,,\\
    \label{eq:trasf1_h03}
    h'_{03}(x^{\prime \rho})=&-d(fh_+ + gh_{\times})\cos(\omega(t'-z'/c))\nonumber\\&-dz'\frac{\omega}{c}(fh_+ + gh_{\times})\sin(\omega(t'-z'/c))\,.
\end{align}
This approach seems promising for meeting our objectives. Still, the diagonal space components of the GW vanish, therefore we have to find the appropriate fluid perturbation such that $\tilde g_{11}$, $\tilde g_{22}$  and $\tilde g_{33}$ vanish or are suppressed. For more suitable gauge transformations see Ref.\,\onlinecite{tesi}.

\subsubsection{Comparison of  GW with the perturbed acoustic metric}
In this section, we  compare the GW metric in the new gauge introduced previously to the acoustic perturbation metric $\tg$ given in Eq.\,\eqref{eq:acousticPerturbMetric}. Recall that to express the background acoustic metric as a Minkowski metric, we assume $v_i=0$ for all $i$ (refer to Eq.\,\eqref{eq:analog_eta_dx0=csdt}).  Now, we aim to identify additional system characteristics that make the acoustic metric perturbation resemble that of a GW. As anticipated in the Introduction, we refer to this perturbation of the acoustic metric, when expressed similarly to a GW in the new gauge, as the GW-like perturbation. We disregard the prefactor of $\tg$, which we consider uniform and time-constant. To align $h'_{\mu\nu}(x^{\prime\rho})$ (the GW metric in the transformed frame) with the acoustic metric $\tg$, we make the following identifications:
\begin{itemize}
    \item The coordinates $t',x',y',z'$ in the new GW reference frame correspond to the time and space coordinates of our BEC system. For simplicity, in the following we drop the primes and refer to these coordinates as $t,x,y,z$ and we also use $h_{\mu\nu}$ instead of $h'_{\mu\nu}$.
    \item The frequency $\omega$ of the GW is identified as the frequency of the GW-like perturbation. 
    \item The speed of light $c$ in the GW metric is replaced by the speed of sound $c_s$ in the analogue system, implying that the GW-like perturbation propagates at $c_s$.
\end{itemize}
Next, let us discuss the  conditions in Eq.\,\eqref{eq:long} in  our physical system. Let $L_x$, $L_y$ and $L_z$ represent the system's dimensions in the $x$, $y$ and $z$ directions, respectively. Defining dimensionless coordinates $\Tilde{x}=x/L_x$, $\Tilde{y}=y/L_y$ and $\Tilde{z}=z/L_z$ we get:
\begin{equation}
    \begin{aligned}
    &x\frac{\omega}{c_s}=\Tilde{x}\frac{L_x}{c_s}\omega=2\pi\Tilde{x}\frac{\omega}{\omega_x},\\
    &y\frac{\omega}{c_s}=\Tilde{y}\frac{L_y}{c_s}\omega=2\pi\Tilde{y}\frac{\omega}{\omega_y},\\
        &z\frac{\omega}{c_s}=\Tilde{z}\frac{L_z}{c_s}\omega=2\pi\Tilde{z}\frac{\omega}{\omega_z},
        \end{aligned}
\end{equation}
where 
\begin{equation}
    \omega_x=\frac{2\pi c_s}{L_x}\,, \quad \omega_y=\frac{2\pi c_s}{L_y}\,, \quad
    \omega_z=\frac{2\pi c_s}{L_z}\,,
\end{equation}
are the three characteristic frequencies.
For a  setting, say an optical trap, that elongates along the $z$--direction, the  transverse frequencies, $\omega_x$ and $\omega_y$ are much larger then the longitudinal frequency, $\omega_z$.  It follows that we can consider perturbation frequencies such that
\begin{equation}
    \label{eq:meaning_approx}
    \frac{x \omega}{c_s}\ll 1 \Leftrightarrow \frac{\omega}{\omega_x}\ll 1, \quad \frac{y \omega}{c_s}\ll 1 \Leftrightarrow \frac{\omega}{\omega_y}\ll 1\\,
\end{equation}
while
\be
\frac{\omega}{\omega_z} \sim 1\,.
\ee
This implies that the considered perturbation in the transverse direction is negligible, while along the longitudinal direction it is comparable with the system  dimension, $L_z$. This is different from Ref.\,\onlinecite{Hartley_GWsimulation} where only GW waveleghts much larger than the BEC size where  considered.\\
At this point, we compare the GW metric $h_{\mu\nu}$ given in Eq.\eqref{eq:hmunu_matrix}, with its components detailed in Eqs. \eqref{eq:trasf1_h00}-\eqref{eq:trasf1_h03}, with
$\tg$ given in Eq.\,\eqref{eq:analog_h_dx0=csdt}. In order to match the two expressions, the space diagonal components $\tilde g_{ii}$, of the acoustic metric perturbation, should vanish. From Eq.\,\eqref{eq:tgig},  we have to impose that the relative density  variation is equal to the relative  scattering length change, that is $\delta a/a=\delta n_c/n_c$.  With this choice, from Eq.\,\eqref{eq:acousticPerturbMetric}, we have that $\tilde g_{00}=-2\delta a/a=-2\delta n_c/n_c$. The space-time dependence of $\delta a/a=\delta n_c/n_c$ is given through the comparison with $h'_{00}$ in Eq.\,\eqref{eq:trasf1_h00}. Then, by equating $h'_{0i}$ in Eqs.\,\eqref{eq:trasf1_h01}-\eqref{eq:trasf1_h03} with the $\tilde g_{0i}$ components in Eq.\,\eqref{eq:analog_h_dx0=csdt}, we find the required space-time dependence of $\delta v_i/c_s$. Combining all these constraints, we conclude that to simulate a GW the system's parameters should  satisfy the following conditions:
\begin{align}
    \label{eq:comparison_trasf1_general}
        \frac{\delta a}{a}=\frac{\delta n_c}{n_c}=&-dz\frac{\omega}{c_s}(f\h_++g \h_\times)\sin(\omega(t-z/c_s))\,,\nonumber\\
        \frac{\delta v_x}{c_s}=&l(f\h_++g\h_\times)\cos(\omega(t-z/c_s))\,,\nonumber\\
        \frac{\delta v_y}{c_s}=&b(f\h_++g\h_\times)\cos(\omega(t-z/c_s))\,,\nonumber\\
        \frac{\delta v_z}{c_s}=&d(f\h_++g\h_\times)\cos(\omega(t-z/c_s))\nonumber\\&+dz\frac{\omega}{c_s}(f\h_++g\h_\times)\sin(\omega(t-z/c_s))\,.
\end{align}
Here, $\h_+$ and $\h_\times$ are the analogue of the GW polarization amplitudes. The first of these equations requires equal spacetime modulations of the scattering length and density perturbations, which is experimentally challenging.   If we set $d=0$, we greatly simply the system, removing all together the need of a Feshbach resonance to vary $\delta a$, of a density fluctuation $\delta n_c$ and of  a velocity fluctuation along the $z$--direction: 
\be
\label{eq:conditions-d=0}
\delta a=\delta n_c=\delta v_z=0\,.
\ee
Under these conditions no modulation of the sound velocity and consequent phonon production is needed. This is different from the case studied in Ref.\,\onlinecite{Hartley_GWsimulation} (see also Ref.\,\onlinecite{Schutzhold}).\\
Equating the GW to the acoustic metric perturbation
\be\label{eq:CC_gauge}
h'_{\mu\nu} \propto\tg     = \frac{n_c}{m c_s }\begin{pmatrix}
    0 & -\frac{\delta v_x}{c_s} &-\frac{\delta v_y}{c_s}&  0\\
   -\frac{\delta v_x}{c_s} &0 &0&0\\
    -\frac{\delta v_y}{c_s} &0&0&0\\
   0 &0&0&0\\
\end{pmatrix}\,,
\ee
we have that \begin{align}
\frac{\delta v_x}{c_s} & =l(f\h_+ +g \h_\times)\cos(\omega(t-z/c_s))\,,\nonumber\\
\frac{\delta v_y}{c_s}&=b(f\h_+ +g\h_\times)\cos(\omega(t-z/c_s))\,.
\label{eq:deltav_d0}
\end{align}
By imposing $d=0$, the GW has only two independent degrees of freedom, see Eqs. \eqref{eq:trasf1_h00}-\eqref{eq:trasf1_h03}, therefore we have completely fixed the gauge. 
This is  determined by
$\zeta_\mu$ in Eqs.\,\eqref{eq:CC_gauge_1}, with $d=0 $. 
This gauge is somehow analogous to the TT gauge, but the only nonvanishing components of the GW are now $h'_{01}$ and $h'_{02}$. 
Since  $\square \zeta_\mu=0$ and $h'_{\mu\nu}$ is traceless, the linearized Einstein's equations take the simple form given in Eq.\,\eqref{eq:Einstein_lin_nosource}.\\
Regarding the metric perturbation in Eq.\eqref{eq:CC_gauge}, it  has been obtained assuming that   the scattering length is constant. Thus,  the velocity fluctuations along the $x$ and $y$ directions should be induced by an additional interaction potential. Such potential will be explored in the next section to determine if a BEC with these characteristics is physically feasible, meaning it satisfies the continuity and Euler equations as well as the irrotational condition.

\subsubsection{Physical systems}
\label{sec:physicalsystems-sec}
We have determined a simple expression of the perturbed acoustic metric that matches the form of a GW in a given gauge.
We need to verify that it can be obtained by appropriate external perturbations of the hydrodynamic quantities, such that the continuity, the Euler equations, as well as the irrotational velocity conditions, are satisfied.\\
In the hydrodynamic limit, the Gross-Pitaevskii equation, see Eq.\,\eqref{eq:GrossQunatumFiledEntire}, can be written as \cite{Barcelò_analoguegravityReview}:
\begin{align}
    \label{eq:continuity_Euler_background}
        \partial_t n_c &=-\nabla \cdot(n_c\mathbf{v})\,,\\
        m\partial_t \mathbf{v}&=-\nabla\left(\frac{m v^2}{2}+V_{\text{ext}}+\frac{4\pi a \hslash^2}{m}n_c\right)\,,
\end{align}
which are the continuity and Euler equation, respectively.
Since we focus on the case where $d=0$, according to Eq.\,\eqref{eq:conditions-d=0} we can restrict to consider  the  deviations from the equilibrium state determined by
\begin{align}
   v_i\to  & \,v_i+\epsilon \,\delta v_i\,,\nonumber\\
    V_{\text{ext}}\to &\, V_{\text{ext}}+\epsilon\,\delta V_{\text{ext}}\,.
\end{align}
Regarding the background, as discussed before, we require $\mathbf{v}=\mathbf{0}$, so that the  unperturbed acoustic metric resembles the Minkowski spacetime. In this case the continuity equation is satisfied if $n_c$ is stationary. The Euler equation is instead satisfied by an appropriate choice of the external potential and scattering length.\\
At leading order in $\epsilon$, the continuity equation holds if $n_c$ is spatially homogeneous. To satisfy the Euler equation with the perturbed velocities in Eq.\,\eqref{eq:deltav_d0}, the external potential perturbation  must be
\begin{equation}
    \delta V_{\text{ext}}=m\omega c_s(lx+by) (f\h_++g\h_\times)\sin(\omega(t-z/c_s))\,.
\label{eq:deltaV}
\end{equation}
Upon inserting such  potential modulation in the Euler equation, one can see that it induces  the velocity  perturbation along the $z$-axis
\begin{equation}
    \label{eq:inducedDELTAVZ1}
     \frac{\delta v_z}{c_s}=\frac{\omega}{c_s}(lx+by)(f \h_++g\h_\times)\sin(\omega(t-z/c_s)\,,
\end{equation}
which is however $\mathcal{O}(\epsilon\omega/\omega_{x,y})$ and thus it is  negligible within our approximation. As discussed before, such $\delta v_z$ is consistent with the velocity irrotational condition: introducing $\delta v_x$ and $\delta v_y$ to simulate a GW naturally induces a negligible velocity fluctuation along the $z$--axis.\\
In summary, we have demonstrated that by applying a coordinate frame transformation from the TT gauge with $\zeta^\mu$ given by Eq.\,\eqref{eq:CC_gauge_1}, the GW metric aligns to the perturbation in the acoustic metric $\tg$. By setting $\mathbf{v}=\mathbf{0}$ the external potential perturbation necessary to have the appropriate velocity perturbations given in  Eq.\,\eqref{eq:deltav_d0} is the one in Eq.\,\eqref{eq:deltaV}.

\section{\label{sec:3}Acoustic BH with a
GW–like perturbation}

So far we have discussed the propagation of a GW-like perturbation in flat spacetime. 
Now, we  consider scenarios  characterized by a different metric. Specifically, we focus on an  acoustic metric  featuring an event horizon and we examine how to appropriately extend a GW-like perturbation in this context. For our analysis, we adopt a cylindrical geometry.\\
Clearly, in this section we are not interested to make  contact with astrophysical objects: both the background metric and the perturbation cannot be mapped to any realistic BH and GW, respectively. Nevertheless, with a minor abuse of language, we will refer to them as the acoustic BH and the GW-like perturbation.

\subsection{Vortex geometry}
\label{sec:ABH_cylindrical}
We consider the `draining bathtub' fluid  model,  representing a $(2+1)$-dimensional flow  with a sink at the origin\,\cite{Visser_acousticBH}. 
For simplicity we neglect any background density variation: the fluid is homogeneous, with constant density, pressure and speed of sound. We use this approximation because we are focusing on a region sufficiently far from the vortex core, where we can neglect the density variation. In this configuration the continuity equation requires that the radial velocity is given by
\begin{equation}
     v_r \propto \frac{1}{r}\,,
\end{equation}
where $r$ is the  distance from the sink. The condition of irrotational (vorticity-free) flow is compatible with the tangential velocity
\be  v_\phi \propto \frac{1}{r}\,,
\ee
thus the fluid velocity can be written as 
\begin{equation}
    \mathbf{v}=\frac{(A \mathbf{\hat{r}}+B\mathbf{\hat{\phi}})}{r}\,,
\label{eq:v_background_cylinder}
\end{equation}
with $A$ and $B$ two constants having dimensions of length$^2$/time. 
The corresponding velocity potential, see Eq.\,\eqref{eq:potential_velocity}, must be expressed as a linear combination of the logarithm of the radial coordinate, $\ln r$, and the angular coordinate $\phi$. Such potential is discontinuous as $\phi$ wraps around by $2\pi$, and hence, it is not monodromic: the velocity potential should be defined patch-wise on overlapping regions surrounding the vortex core at $r=0$. Neglecting any overall space-independent prefactor, the acoustic metric for this `draining bathtub' scenario is
\begin{equation}
    \mathrm{d}s^2=-c_s^2\mathrm{d}t^2+\left( \mathrm{d}r-\frac{A}{r}\mathrm{d}t \right)^2+\left( r \mathrm{d}\phi-\frac{B}{r}\mathrm{d} t\right)^2\,,
\end{equation}
and an acoustic event horizon forms at the radial distance
\begin{equation}
    \label{eq:acHOR_posit_vortex}
    r_H=\frac{|A|}{c_s}\,,
\end{equation}
where the radial  fluid velocity is equal to the speed of sound. The sign of $A$ determines the nature of the horizon: we take  $A<0$, thus the horizon is a future acoustic horizon (acoustic BH), whereas for $A>0$, corresponds to  a past acoustic horizon (acoustic white hole).
\begin{figure}[!b]
    \centering
\includegraphics[width=0.25\textwidth]{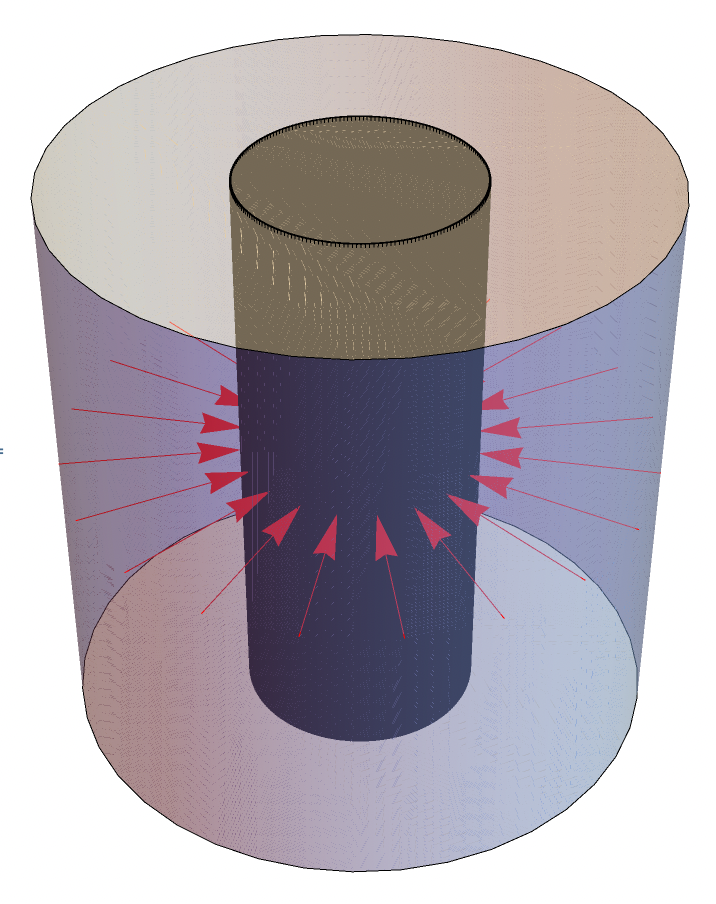}
    \caption{Pictorial  representation of a GW-like perturbation in the draining bathtub geometry.    The dark cylinder represents the event horizon, while pink arrows indicate the  propagation direction  of the GW-like perturbation.}
    \label{fig:ABH+GW_schematic}
\end{figure}
Although this construction is described in $(2+1)$--dimensions, it can be extended to $(3+1)$--dimensions by adding an extra spatial dimension, $z$, indicating the height of the cylinder. 
The condition of irrotational flow implies that the $z$--component of the velocity is constant; for simplicity we assume that it  vanishes. Hence in  $(3+1)$--dimensions we have a vortex filament located at  $r=0$ generating the  metric 
\begin{equation}
    \label{eq:cylindricalBH_metric}
    \mathrm{d} s^2=-c_s^2 \mathrm{d}t^2+\left( \mathrm{d}r-\frac{A}{r}\mathrm{d} t \right)^2+\left(r \mathrm{d}\phi -\frac{B}{r}\mathrm{d}t\right)^2+\mathrm{d}z^2\,,
\end{equation}
for the phonon excitations. This extended metric will serve as the foundation for our exploration of a system where an acoustic BH is perturbed by a GW-like perturbation that propagates radially. A schematic representation of the system is shown in Fig.\,\ref{fig:ABH+GW_schematic}. This choice of geometry is entirely distinct from astrophysical BHs. The background metric given by Eq.\eqref{eq:cylindricalBH_metric}, with $A<0$, has indeed a cylindrical  event horizon  located at $r_H$ for all $z$. We have chosen this specific  geometry for several reasons. While simpler configurations, such as a two-dimensional horizon, could be considered, the cylindrical case offers a more general approach and holds significant experimental potential. Laboratory experiments have successfully created box-like BECs \cite{BECinaBox, HadzibabicBOX, ZwierleinBOX}, suggesting the feasibility of confining a BEC within a cylindrically symmetric trapping potential. This geometry can thus serve as a practical model for experimental analogues of gravitational phenomena. Moreover, cigar-shaped or disk-shaped BECs, which have been used in analogue gravity experiments\cite{2019MunozdeNova, 2019HuChin}, can be obtained as  limiting cases of  the cylindrical BEC.

\subsection{Method}

We proceed to consider the various steps needed to realize a GW-like perturbation in the cylindrical geometry:
\begin{enumerate}
    \item \textit{Cylindrical perturbations}: Extend the externally driven GW-like perturbation obtained in Sec.\,\ref{sec:GW_flat} to the cylindrical geometry.
    \item \textit{Physical requirements}: Check the irrotational condition constraint, as well as the continuity and Euler equations.
    \item \textit{Acoustic metric}: Calculate the acoustic metric and its inverse to characterize the emergent spacetime.
    \item \textit{Perturbed acoustic horizon}: Investigate the position of the acoustic horizon upon interaction with the analogue GW.
\end{enumerate}
As in Sec.\,\ref{sec:GW_flat}, it is important to distinguish between the externally driven metric perturbations and phonons. In Appendix \ref{app:A}, the generators of the acoustic horizon are examined to elucidate their properties under perturbations.

\subsubsection{Cylindrical perturbations}
We extend the GW-like plane wave perturbation obtained in Sec.\,\ref{sec:GW_flat} to a cylindrical geometry. In the linear case discussed earlier, we identified the analogue of a gravitational wave by perturbing the condensate velocity along the directions perpendicular to the wave's propagation. For the cylindrical geometry, we consider that the perturbation propagates radially, as depicted by pink arrows in Fig.\,\ref{fig:ABH+GW_schematic}. Consequently, the GW-like perturbation  is characterized by velocity fluctuations along the angular and $z$--coordinates, denoted as $\delta v_\phi$ and $\delta v_z$, respectively. These perturbations can be obtained from those in Eq.\,\eqref{eq:deltav_d0} with the replacements  $\delta v_x \to \delta v_z$ and $\delta v_y \to r/r_0 \delta v_\phi$, where $r_0$ is a normalizing  length. Due to the radial propagation direction, the term $\cos(\omega(t-z/c_s)$ in Eq.\,\eqref{eq:deltav_d0} is replaced by $\cos(\omega(t-r/c_s))$. In conclusion, the GW-like perturbation in the cylindrical case is identified by the following velocity fluctuations:
\begin{align}
    \frac{\delta v_z}{c_s}=&\,l(f \h_++g\h_\times)\cos(\omega(t-r/c_s)),\,\label{eq:deltav_cylinder1}\\
    \frac{\delta v_\phi}{c_s}=&\frac{r_0}{r}b(f\h_++g\h_\times)\cos(\omega(t-r/c_s)).
\label{eq:deltav_cylinder2}
\end{align}
The characteristic frequencies of this system are
\begin{equation}
    \label{eq:characteristic_frequencies}
    \omega_z=\frac{2\pi c_s}{L_z}, \quad    \omega_r=\frac{2\pi c_s}{L_r},
\end{equation}
where $L_z$ and $L_r$ represent the physical dimensions of the system along the $z$ and $r$ directions, respectively.
Given that the system elongates along the $z$--direction, we assume that 
\be
\omega_r > \omega_z\,,
\ee
and we consider perturbation frequencies 
\be\label{eq:frequency_cylinder}
\omega \sim \omega_z\,,
\ee
thus with a wavelength of order  the height of the cylinder.

\subsubsection{Physical requirements}
We now check whether the irrotational condition, as well as the continuity, and Euler equations hold for the velocity perturbations in Eqs.\,\eqref{eq:deltav_cylinder1} and \eqref{eq:deltav_cylinder2}. As in the Minkowskian case, we assume that the unperturbed system is homogeneous:  the  density, pressure and  the speed of sound are taken constant. The unperturbed velocity is given by Eq.\,\eqref{eq:v_background_cylinder} with $A$ and $B$ two constants. \\
We begin by examining the irrotationality of the flow, checking that $\nabla \times \delta \mathbf{v}=\mathbf{0}$. 
We have seen that in flat spacetime the irrotational condition requires the presence of an additional velocity component; likewise here the fluctuations $\delta v_z$ and $\delta v_\phi$ satisfy the irrotational condition only if 
\begin{align}
\delta v_r  =&lz\omega (f\h_++g\h_\times)\sin(\omega(t-r/c_s))\nonumber\\
&+b \phi r_0 \omega (f\h_++g\h_\times)\sin(\omega(t-r/c_s)) \,,
\end{align}
which  is thus the sum of two terms: one proportional to $l\,z$ and the other one to $b\,\phi$. In order to have a single-valued function we impose this second term to be zero taking  $b=0$. This constraint, in turn, implies that $\delta v_\phi=0$, see Eq.\,\eqref{eq:deltav_cylinder2}. Hence, we start with a GW-like perturbation that is represented only by the velocity perturbation $\delta v_z$ in Eq.\,\eqref{eq:deltav_cylinder1} and this induces a radial velocity perturbation 
\begin{equation}
\label{eq:deltavr_cyl}
    \frac{\delta v_r}{c_s}=lz\frac{\omega}{c_s}(f\h_++g\h_\times)\sin(\omega(t-r/c_s)).
\end{equation}
In the Minkowskian case,  the induced velocity perturbation was negligible (see Eq. \,\eqref{eq:inducedDELTAVZ1}), because it was of order ${\cal O }(\epsilon \omega/\omega_{x,y})$. In  cylindrical geometry it is instead ${\cal O }(\epsilon \omega/\omega_{z})$   and we cannot neglect it, see Eq.\,\eqref{eq:frequency_cylinder}. Note that  the  radial velocity perturbation depends on the $z$ coordinate, hence it produces a time dependent bending of the acoustic horizon that changes with height. 

In order to simplify the analysis, we turn to dimensionless coordinates: we express $z$ as $\Tilde{z}L_z$ with $\bar{z}\in[0,1]$ and $r$ as $\bar{r}L_r$ with $\bar{r}\in[0,1]$. For convenience we choose $l=1/2\pi$ and so:
\begin{align} 
    &\frac{\delta v_z}{c_s}=\frac{1}{2\pi}(f\h_++g\h_\times)\cos(\omega(t-2\pi \bar{r}/\omega_r)),\nonumber\\
    &\frac{\delta v_r}{c_s}= \bar{z}\frac{\omega}{\omega_z}(f\h_++g\h_\times)\sin(\omega(t-2\pi \bar{r}/\omega_r)).
    \label{eq:deltav_BH}
\end{align}
In the Minkowskian case, the velocity perturbations in Eq.\,\eqref{eq:deltav_d0} were solutions of the continuity and Euler equations with $\delta a=\delta n_c=0$. However, in the cylindrical case with an acoustic horizon, the velocity perturbations in Eq.\,\eqref{eq:deltav_BH} remain solutions of these equations, provided we introduce appropriate additional terms: the continuity equation forces us to take $\delta n_c\not=0$, and as a consequence $\delta a\not=0$. Specifically, the continuity equation holds if we include the source term 
\begin{align}
        \label{eq:S_pert}
    \delta S  =&\frac{2\pi}{3}n_c\frac{\bar{z} \bar{r}^3}{\bar{r}_H^2}\frac{\omega^3}{\omega_r \omega_z}(f\h_++g\h_\times)\nonumber\\ &\times\left(-1-\frac{\bar{r}_H}{\bar{r}}\right)\sin(\omega(t-2\pi \bar{r}/\omega_r))\,,
\end{align}
where 
\begin{equation}\label{eq:rh_nodim}
    \bar{r}_H=\frac{|A|}{c_s L_r}\,,
\end{equation}
is the unperturbed horizon position, see Eq.\,\eqref{eq:acHOR_posit_vortex}, in dimensionless units. Such source term induces the density variation
\begin{align}
    \label{eq:n_c_pert}
    \frac{\delta n_c}{n_c}= &\frac{\bar{z}\bar{r}}{\bar{r}_H}\frac{\omega}{\omega_z}(f\h_++g\h_\times)\sin(\omega(t-2\pi \bar{r}/\omega_r))\nonumber\\ &+\frac{2\pi}{3}\frac{\bar{z}\bar{r}^3}{\bar{r}_H^2}\frac{\omega^2}{\omega_r \omega_z}(f\h_++g\h_\times)\cos(\omega(t-2\pi \bar{r}/\omega_r)).
\end{align}
Furthermore, to maintain $c_s$ as the local speed of sound, we require $\delta c_s=0$, which leads to the introduction of a Feshbach resonance (see Eq.\,\eqref{eq:c_s}) such that $\delta a/a=-\delta n_c/n_c$. With this choice of $\delta a$, the Euler equation holds if we include the external potential \begin{align}
\label{eq:Vext_pert}
    \delta V_{\text{ext}}= &\left(\frac{3}{8\pi}\frac{\bar{z}(\bar{r}-\bar{r}_H)}{\bar{r}_H^2}\frac{\omega^2 \omega_r}{\omega_z c_s^2}-\frac{\pi}{6}\frac{\bar{z}\bar{r}^3}{\bar{r}_H^2}\frac{\omega^4}{\omega_z \omega_r c_s^2}\right)\frac{\hslash^2}{m}\nonumber\\
    &\times(f\h_++g\h_\times)\cos(\omega(t-2\pi\bar{r}/\omega_r))-\left(\bar{z}\frac{c_s^2\omega}{\omega_z}m  \right.\nonumber\\
    &\left.+\frac{1}{16\pi^2}\frac{\bar{z}}{\bar{r}\bar{r}_H}\frac{\omega \omega_r^2}{\omega_z c_s^2}\frac{\hslash^2}{m}+\frac{\bar{z}\bar{r}}{\bar{r}_H^2}\left(\frac{7}{12}\bar{r}-\frac{1}{4}\bar{r}_H\right)\frac{\omega^3}{\omega_z c_s^2}\frac{\hslash^2}{m}\right)\nonumber\\
    &\times(f\h_++g\h_\times)\sin(\omega(t-2\pi\bar{r}/\omega_r)).
\end{align}
In this way we obtain a physical system (irrotationality, continuity and Euler equations are satisfied), in the geometry depicted in Fig.\,\ref{fig:ABH+GW_schematic}, that represents an acoustic BH with a GW-like perturbation.

\subsubsection{Acoustic metric}
\label{sec:acousticmetricBHGW}
At this point, we compute the emergent acoustic metric tensor for phonons, given by $g_{\mu\nu}=g^0_{\mu\nu}+\epsilon\tg$.
For simplicity we assume vanishing tangential flow, that is we take   $B=0$ in Eq.\,\eqref{eq:v_background_cylinder}. Using coordinates $(c_s t, \bar{r}, \theta, \bar{z})$, the unperturbed metric is
\begin{equation}
    \label{eq:eta_coordTILde}
    g^0_{\mu\nu}=\frac{n_c}{m c_s}\begin{pmatrix}
        -\left(1-\frac{v_r^2}{c_s^2}\right)&-\frac{v_r}{c_s}L_r&0&0\\
        -\frac{v_r}{c_s}L_r& L_r^2&0&0\\
        0&0&\bar{r}^2L_r^2&0\\
        0&0&0&L_z^2\\
    \end{pmatrix}
    ,
\end{equation}
and the perturbation 
\begin{equation}
    \label{eq:h_coordTILde}
    \tg=\frac{n_c}{m c_s}\begin{pmatrix}
       \tilde g_{00}&\tilde g_{01}&0&\tilde g_{03}\\
       \tilde g_{01}&\tilde g_{11}&0&0\\
        0&0&\tilde g_{22}&0\\
        \tilde g_{03}&0&0&\tilde g_{33}\\
    \end{pmatrix}
    ,
\end{equation}
has components 
\begin{equation}
     \tilde g_{00}=-\left(1-\frac{v_r^2}{c_s^2}\right)\frac{\delta n_c}{n_c}+2\frac{v_r\delta v_r}{c_s^2},
\end{equation}
\begin{equation}
    \tilde g_{01}=L_r\left(-\frac{\delta v_r}{c_s}-\frac{v_r}{c_s}\frac{\delta n_c}{n_c}\right),
\end{equation}
\begin{equation}
    \tilde g_{03}=-\frac{\delta v_z}{c_s}L_z\,, \quad
    \tilde g_{11}=\frac{\delta n_c}{n_c}L_r^2,
\end{equation}
\begin{equation}
    \tilde g_{22}=\Tilde{r}^2 L_r^2\frac{\delta n_c}{n_c}\,, \quad
    \tilde g_{33}=\frac{\delta n_c}{n_c}L_z^2\,,
\end{equation}
that are the variations of the acoustic metric induced by the GW-like perturbation. Far from the horizon, they can be seen as  perturbations of the Minkowski metric. Near the horizon, they represent a  perturbation of the cylindrical acoustic metric.
Next, we compute the inverse  of the metric defined by the relation $g_{\mu\nu}g^{\nu\alpha}=\delta_{\mu}{}^{\alpha}$, and  we write it as
\begin{equation}
    \label{eq:gnualpha2}
    g^{\mu\nu}=g^{\mu\nu}_0-\epsilon \tilde g^{\mu\nu}.
\end{equation}
In the coordinates $(c_s t, \Tilde{r}, \theta, \Tilde{z})$, the unperturbed inverse metric is
\begin{equation}
    \label{eq:etaINV_coordTilde}
    g^{\mu\nu}_0=\frac{c_s m}{n_c}\begin{pmatrix}
        -1&-\frac{v_r}{c_s L_r}&0&0\\
        -\frac{v_r}{c_s L_r}&\frac{1}{L_r^2}\left(1-\frac{v_r^2}{c_s^2}\right)&0&0\\
        0&0&\frac{1}{\Tilde{r}^2L_r^2}&0\\
        0&0&0&\frac{1}{L_z^2}\\
    \end{pmatrix}
    ,
\end{equation}
and the inverse perturbation  is
\begin{equation}
    \label{eq:hINV_coordTilde}
    \tilde g^{\mu\nu}=\frac{mc_s}{n_c}\begin{pmatrix}
        \tilde g^{00}&\tilde g^{01}&0&\tilde g^{03}\\
        \tilde g^{01}&\tilde g^{11}&0&\tilde g^{13}\\
        0&0&\tilde g^{22}&0\\
        \tilde g^{03}&\tilde g^{13}&0&\tilde g^{33}
    \end{pmatrix}
    ,
\end{equation}
with  components
\begin{equation}
    \tilde g^{00}=-\frac{\delta n_c}{n_c},
\end{equation}
\begin{equation}
    \tilde g^{01}=\left(-\frac{\delta n_c}{n_c}\frac{v_r}{c_s}+\frac{\delta v_r}{c_s}\right)\frac{1}{L_r},
\end{equation}
\begin{equation}
    \tilde g^{03}=\frac{\delta v_z}{c_s}\frac{1}{L_z},
\end{equation}
\begin{equation}
    \tilde g^{11}=\left(2\frac{v_r \delta v_r}{c_s^2}+\frac{\delta n_c}{n_c}-\frac{v_r^2}{c_s^2}\frac{\delta n_c}{n_c}\right)\frac{1}{L_r^2},
\end{equation}
\begin{equation}
    \tilde g^{13}=\frac{v_r\delta v_z}{c_s^2}\frac{1}{L_r L_z},
\end{equation}
\begin{equation}
    \tilde g^{22}=\frac{1}{\Tilde{r}^2 L_r^2}\frac{\delta n_c}{n_c},
\end{equation}
\begin{equation}
    \tilde g^{33}=\frac{\delta n_c}{n_c}\frac{1}{L_z^2}.
\end{equation}
From the inverse  metric
we can   determine the event horizon's characteristics and generators.

\subsubsection{Perturbed acoustic horizon}
\label{sec:perturbedAH}
\begin{figure}[!htbp]
    \centering\includegraphics[width=0.45\textwidth]{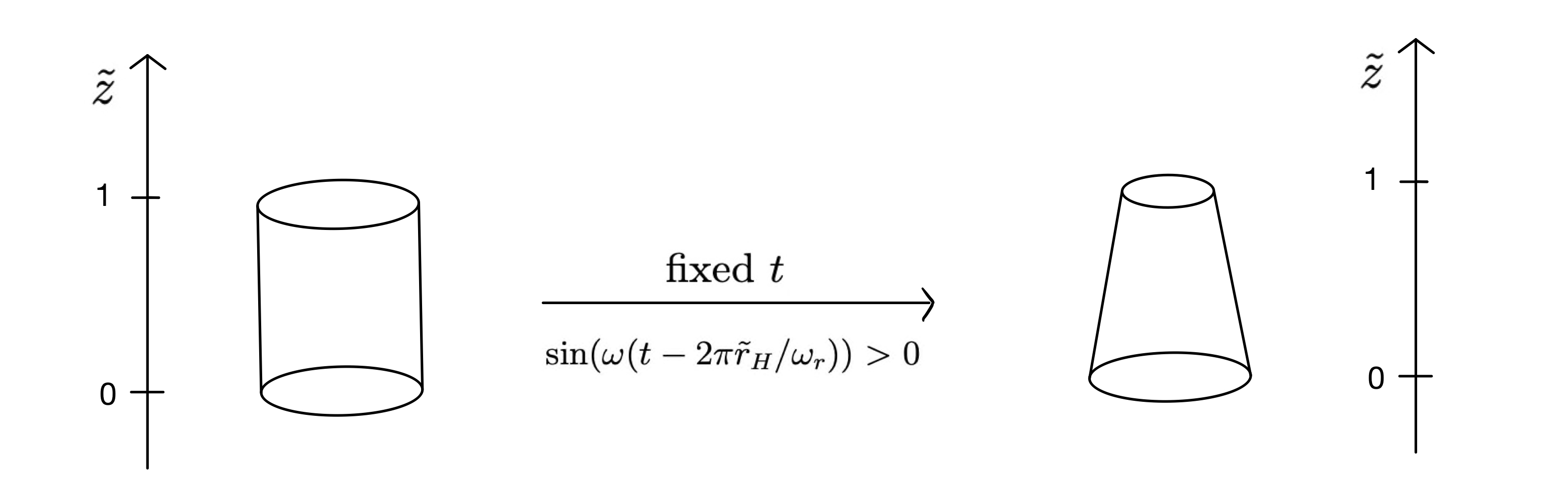}
    \caption{Schematic representation of the perturbation of the acoustic horizon. Left: unperturbed horizon. Right: perturbed acoustic horizon at a fixed time such that $\sin(\omega(t-2\pi\Tilde{r}_H/\omega_r))>0$. The size of the perturbation is enhanced for clarity.}
    \label{fig:deltarh_tfix_sin>0}
\end{figure}

\begin{figure}[!htbp]
    \centering
    \includegraphics[width=0.45\textwidth]{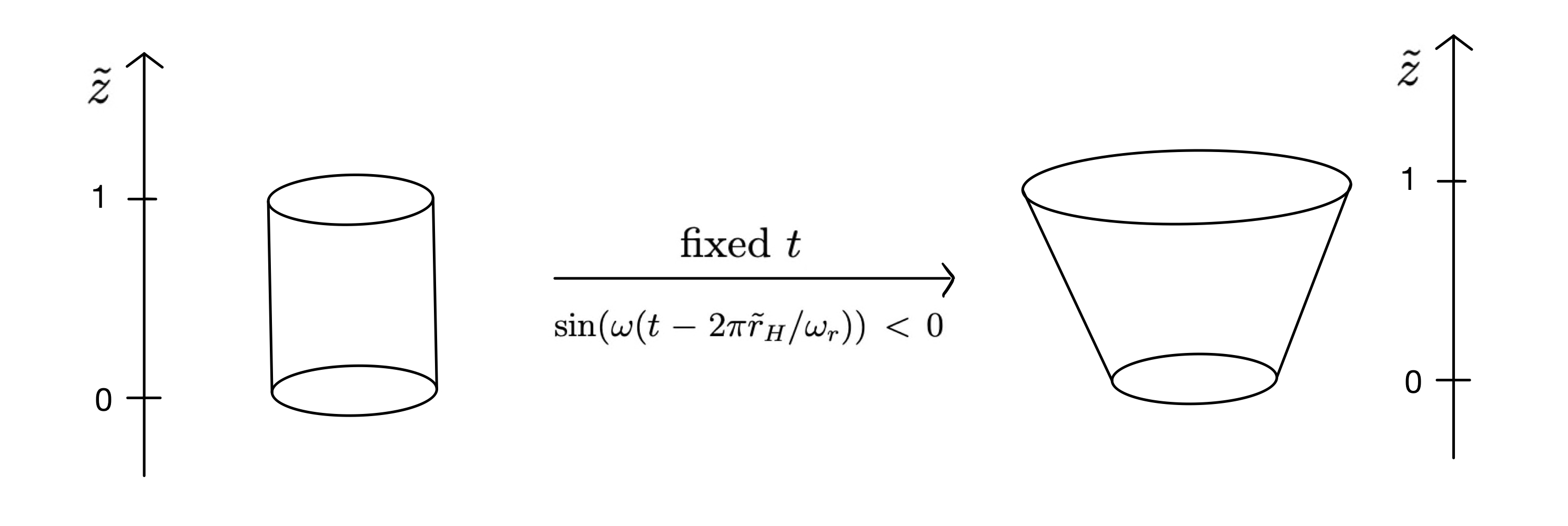}
    \caption{Schematic representation of the perturbation of the acoustic horizon. Left: unperturbed horizon. Right: perturbed acoustic horizon at a fixed time such that  $\sin(\omega(t-2\pi\Tilde{r}_H/\omega_r))<0$. The size of the perturbation is  enhanced for clarity.}
    \label{fig:deltarh_tfix_sin<0}
\end{figure}

\begin{figure}[!htbp]
    \centering
    \includegraphics[width=0.45\textwidth]{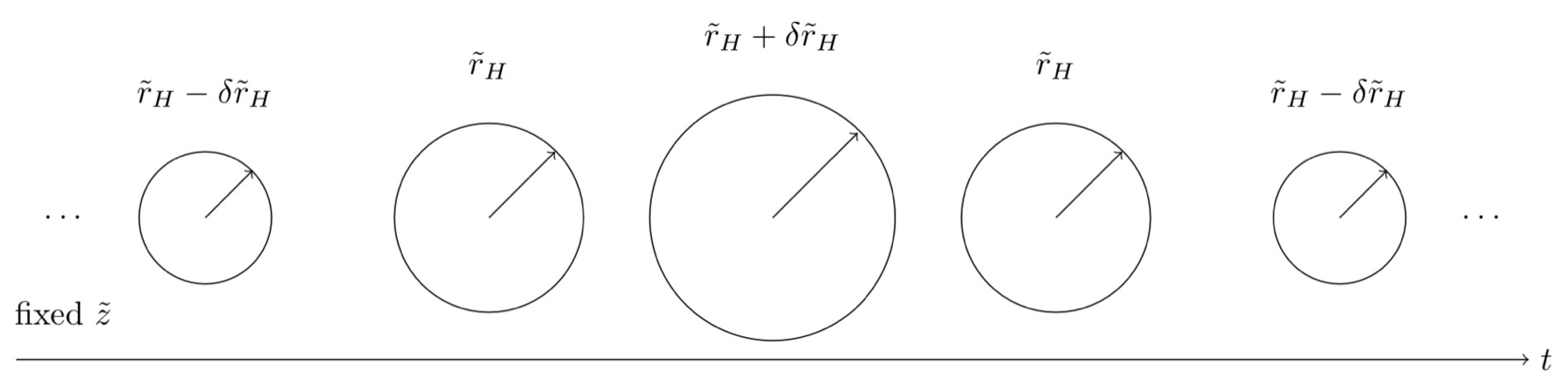}
    \caption{Schematic representation of the perturbation of the perturbed acoustic horizon in the $\Tilde{x}\Tilde{y}$ plane. At fixed $\Tilde{z}>0$ the horizon is a circle that oscillates as time lapses.}
    \label{fig:deltarh_zfix}
\end{figure}

We now determine the displacement of the horizon position determined by the impinging GW-like perturbation. In general, the horizon position is given by the condition $|v_r|_{r_{H}}=c_s$. In the unperturbed system, the horizon is located in $\bar r_H$ given in Eq.\,\eqref{eq:rh_nodim}.
The GW-like propagation is determined by the velocity fluctuation $\delta v_r$ in Eq.\,\eqref{eq:deltavr_cyl}. Consequently, the acoustic horizon is displaced at $\Tilde{r}_H^{\text{new}}=\Tilde{r}_H+\epsilon \delta \Tilde{r}_H$, which satisfies:
\begin{equation}
    \label{eq:newhor_eq}
    |v_r+ \epsilon\delta v_r|_{\Tilde{r}_H^{\text{new}}}=c_s.
\end{equation}
Solving Eq.\,\eqref{eq:newhor_eq}, we find the horizon's displacement is
\begin{equation}
    \label{eq:deltaRH}
    \frac{\delta \Tilde{r}_H}{\Tilde{r}_H}=-\Tilde{z}\frac{\omega}{\omega_z}(f\h_++g\h_\times)\sin(\omega(t-2\pi\Tilde{r}_H/\omega_r))\,,
\end{equation}
indicating that it is opposite to the sign of the radial velocity perturbation, see Eq.\,\eqref{eq:deltav_BH}, while $\delta v_z$ has no effect on the horizon's position.
The horizon's displacement vanishes at $\Tilde{z}=0$ and for $\Tilde{z}\neq 0$ it depends on the sign of $\sin(\omega(t-2\pi\Tilde{r}_H/\omega_r))$.
At a given time $t$, such that $\sin(\omega(t-2\pi\Tilde{r}_H/\omega_r))>0$, we have that $\delta\Tilde{r}_H<0$  and the displacement decreases as $\Tilde{z}$ increases, see Fig.\, \ref{fig:deltarh_tfix_sin>0}.
At time $t$ such that $\sin(\omega(t-2\pi\Tilde{r}_H/\omega_r))<0$, we have he opposite behavior: $\delta \Tilde{r}_H$ increases linearly with $\Tilde{z}$, see Fig.\,\ref{fig:deltarh_tfix_sin<0}.
If we now examine the scenario at fixed $\Tilde{z}\neq 0$, and we observe the changes over time, we find that the horizon displacement fluctuates around zero as shown in Fig. \ref{fig:deltarh_zfix}: the event horizon expands or contracts due to the GW-like  propagation. The magnitude of the dilatation depends on the considered height: it vanishes for $\bar z=0$ and it is maximal for $\bar z=1$. 
Differently from the flat spacetime case where GWs do not cause phonon production, when GWs impinge onto an acoustic horizon they induce on it periodic deformations which turns into a stimulated amplification of the Hawking phonon radiation in a phenomenon resembling the dynamical Casimir effect \cite{Sabin:2014bua}. This process was described in details in Ref.\,\onlinecite{CGMTCarXiv} in the case of a planar acoustic horizon under the action of a small velocity shear in background flow. Something similar takes place for the system considered here with a cylindrical geometry and the shear perturbation being originated by the impinging GWs. The expansion or contraction observed at a fixed $\Tilde{z}$ corresponds to an oscillation of the tilting of the horizon. Due to this tilt, the radiated phonons, which are predominantly emitted in the direction perpendicular to the horizon, experience a change in their propagation direction as compared to the unperturbed case\cite{1983.book.Shapiro, Mannarelli:emissione}. Focusing for instance on the $\Tilde{x}\Tilde{z}$ plane at $\Tilde{y}=0$ and considering only the half-plane $\Tilde{x}>0$, the situation closely resembles the analysis conducted in Sec. V of Ref.\,\onlinecite{CGMTCarXiv}, where the shear viscosity to entropy density ratio $\eta/s$ at the acoustic horizon was calculated to be equal to $1/4\pi$, namely, the KSS lower bound\,\cite{Kovtun:2004de}.
Although a planar horizon was considered in Ref.\,\onlinecite{CGMTCarXiv}, that result holds here as well: in both cases the relevant effect is that given by the bending of the horizon surface with respect to the $\Tilde{z}$ axis.

\section{\label{sec:conclusions}Conclusions and future perspectives}

Developing a system that enables an in-depth investigation of acoustic horizons through perturbations is crucial for addressing fundamental theoretical questions regarding the physics of BHs. In this paper, we have designed a perturbation that closely mimics a classical GW and applied it to excite an acoustic horizon. We first demonstrate that it is possible to configure a BEC such that the metric perturbation experienced by phonons resembles that of a GW in Minkowski space in a specific gauge. Subsequently, we considered the effect of this perturbation, adapted to the new symmetry of the system, on an acoustic cylindrical BH. With the proposed  approach we have  established a laboratory-reproducible system where to study how an acoustic horizon responds to perturbations  closely modeled after GWs. In particular, experiments with ultra-cold atoms can be designed to replicate these systems in order to test the theoretical predictions arising from this research. These models open avenues for diverse future studies.\\
Firstly, the shear viscosity to entropy density ratio of the acoustic horizon can be assessed. One approach is to determine and compare its value with that of a real BH in general relativity \cite{ChircoLiberati_noneqThermo}. In this case, it is important to check whether this ratio adheres the universal lower bound of $\eta/s$, proposed in the AdS/CFT context\cite{Kovtun:2004de}. This bound is conjectured to be $1/4\pi$ for any fluid in nature \cite{overviewETA/S, KSS_holography-hydrodynamics}. However, the conditions under which this bound can be saturated by a matter system, given the known symmetries, remain an open question \cite{2012:eta/sMatter}. Thus, investigating $\eta/s$ for acoustic BHs could be crucial to determine if it saturates this bound and to establish whether the ratio for acoustic BHs is universal or dependent on specific hydrodynamic conditions.\\
Additionally, the surface reflectivity of the acoustic horizon can be explored and related to the membrane fluid viscosity \cite{2020Oshita}. \\
Moreover, the GW-like perturbation discussed in this paper can be analyzed in greater details. A particularly intriguing aspect is how, despite the absence of spin--2 modes in the system, we can still replicate a spin--2 perturbation of the metric. It might be worthwhile to study the quantization of this classical analogue of GWs to investigate whether its quanta can be identified and to predict the equation of motion for this mode. Although quasi-normal modes of acoustic BHs have been studied previously \cite{2004QNM}, examining this spin--2 perturbation could potentially provide a new avenue for approximating a spin--2 test field analysis \cite{testfieldApprox}.
Furthermore, we can explore gravitational memory in these systems to understand how the emergent spacetime responds when a GW-like perturbation passes through it, and whether there are any symmetries of the system related to it\cite{2010GWmemory}.


\begin{acknowledgments}
M.L.C. acknowledges support from the National Centre on HPC, Big Data and Quantum Computing-SPOKE 10 Code CN00000013 (Quantum Computing) and received funding from the European Union Next-GenerationEU-National Recovery and Resilience Plan (NRRP)-MISSION 4 COMPONENT 2, INVESTMENT N. 1.4-CUP N. I53C22000690001. This research has received funding from the European Union’s Digital Europe Programme DIGIQ under grant agreement no. 101084035. M.L.C. also acknowledges support from the project PRA\_2022\_2023\_98 “IMAGINATION”, and in part by grants NSF PHY-1748958 and PHY-2309135 to the Kavli Institute for Theoretical Physics (KITP). D.G. presently on leave of absence at Embassy of Italy, The Hague.
\end{acknowledgments}

\section*{Author declaration}
The authors have no conflicts to disclose.

\section*{Data availability}
Data sharing is not applicable to this article as no new data were created or analyzed in this study.

\appendix

\section{\label{app:A} Horizon's generators}
\label{sec:horizon_generators-sec}
To characterize the kinematics of the horizon, we compute the null vector field that is simultaneously tangent and orthogonal to the null congruence of the geodesics generating the horizon. The vector field normal to the hypersurface $S=\text{constant}$ has the general expression
\begin{equation}
    k^\mu=\ell(x)g^{\mu\nu}\partial_\nu S\,,
\end{equation}
with $\ell(x)$ an arbitrary non-vanishing function of the coordinates. Hereafter we take for simplicity $\ell(x)=1$. \\
We first compute $k^\mu$  in the  unperturbed system. We consider the family of hypersurfaces  
\begin{equation}
    S=\Tilde{r}-\Tilde{r}_H=\text{constant}\,,
\end{equation}
where $\bar{r}_H$ is given in Eq.\,\eqref{eq:rh_nodim}; the acoustic horizon corresponds to $S=0$. For the considered hypersurface
 \be
 \partial_\nu S =(0,1,0,0)\,,
 \ee
hence, using the inverse of the metric in Eq.\,\eqref{eq:etaINV_coordTilde}, we obtain that the vector normal to the hypersurface $S$ is given by
\begin{equation}
    \label{eq:unperturbedKmu}
    k^\mu=\left(\frac{m}{2\pi n_c}\omega_r \frac{\tilde r_H}{\tilde r},\frac{m}{4\pi^2 n_c c_s}\omega_r^2\left(1-\frac{\Tilde{r}_H^2}{\Tilde{r}^2}\right),0,0\right),
\end{equation}
and then
\begin{equation}
    g^0_{\mu\nu}k^\mu k^\nu =\frac{m}{4 \pi^2 n_c c_s}\omega_r^2\left(1-\frac{\tilde r_H^2}{\tilde r^2}\right).
\end{equation}
Since at the horizon $v_r^2=c_s^2$, we find that 
\begin{equation}
\left.    g^0_{\mu\nu}k^\mu k^\nu \right|_{S=0}=0\,,
\label{eq:k2unperturbed}
\end{equation}
meaning  that the acoustic horizon is a null hypersurface.\\
For the computation of the null vector field in the system perturbed by the GW-like perturbation,  we have to consider that the inverse of the metric is given by
Eq.\,\eqref{eq:gnualpha2} and that 
 the position of the  acoustic horizon  is shifted  to $\Tilde{r}_H +\epsilon\delta \Tilde{r}_H$, where $\delta \Tilde{r}_H$ is  in Eq.\,\eqref{eq:deltaRH}. For these reasons we define a new  family of hypersurfaces:
\begin{equation}
    S_\text{new}=\Tilde{r}-\left(\Tilde{r}_H+\epsilon\delta \Tilde{r}_H\right)=\text{constant}\,,
\end{equation}
and the acoustic horizon hypersurface is  $S_\text{new}=0$. 
In this case
 \be
 \partial_\nu S_\text{new} =(0,1,0,0) - \epsilon \partial_\nu \delta \Tilde{r}_H \,,
 \ee
thus the components of the normal vector field, that we now call $k_\text{new}^\mu$, at the leading order in the perturbations are given by 
\be
k_\text{new}^\mu = g_0^{\mu\nu} \partial_\nu S -\epsilon  \tilde g^{\mu\nu} \partial_\nu S - \epsilon g_0^{\mu\nu} \partial_\nu \delta \Tilde{r}_H\,.
\ee
Therefore,  $k_\text{new}^\mu$ is identical to that of the unperturbed system at zeroth order in $\epsilon$, see Eq.\,\eqref{eq:unperturbedKmu}, with corrections at order $\epsilon$ arising from both metric perturbations and changes in the horizon's position. Its components are given by
\begin{eqnarray}
    k_\text{new}^0&&=k^0+\epsilon(f\h_++g\h_\times)\left[\frac{m}{n_c}\frac{\Tilde{z}}{\pi}\frac{\omega \omega_r}{\omega_z}\sin(\omega T_\text{eff})
    \right.\nonumber\\
    &&\left.-\left(\frac{1}{3}\frac{m}{n_c}\frac{\omega^2}{\omega_z}\frac{\Tilde{r}^2\Tilde{z}}{\Tilde{r}_H}+\frac{m}{n_c}\Tilde{z}\Tilde{r}_H \frac{\omega^2}{\omega_z}\right)\cos(\omega T_\text{eff})\right],
\end{eqnarray}
\begin{eqnarray}
    k_\text{new}^1&&=k^1+\epsilon (f\h_++g\h_\times)\left[\left(\frac{m}{2\pi n_c c_s}\frac{\omega^2\omega_r}{\omega_z}\frac{\Tilde{z}\Tilde{r}_H^2}{\Tilde{r}}\right.\right.\nonumber\\
    &&
    \left.+\frac{m}{6\pi n_c c_s}
    \frac{\omega^2\omega_r}{\omega_z}\frac{\Tilde{z}\Tilde{r}}{\Tilde{r}_H^2}(-\Tilde{r}^2+\Tilde{r}_H^2)\right)\cos(\omega T_\text{eff})\nonumber\\
    &&
    \left.
    -\frac{m}{4\pi^2 n_c c_s}\frac{\omega\omega_r^2}{\omega_z}\frac{\Tilde{z}}{\Tilde{r}\Tilde{r}_H}(\Tilde{r}^2-3\Tilde{r}_H^2)\sin(\omega T_\text{eff})\right],
\end{eqnarray}
\begin{equation}
    k_\text{new}^2=0,
\end{equation}
\begin{eqnarray}
    k_\text{new}^3&&=\epsilon(f\h_+ +g\h_\times)\left(\frac{m}{4\pi^2 n_c c_s}\omega\omega_z\Tilde{r}_H \sin(\omega T_\text{eff}) \right.\nonumber\\
    &&\left. +\frac{m}{8 \pi^3 n_c c_s}\omega_z\omega_r\frac{\Tilde{r}_H}{\Tilde{r}} \cos(\omega T_\text{eff})\right),
\end{eqnarray}
where for simplicity we have defined $T_\text{eff}=t-2\pi\Tilde{r}/\omega_r$.
Now, we compute $g_{\mu\nu}k_\text{new}^\nu k_\text{new}^\mu$ at the acoustic horizon, so at $S_\text{new}=0$, and we call it $K$. We find:
\begin{equation}
    K=\epsilon 4\pi \frac{n_c c_s}{m}\frac{\omega^2}{\omega_r \omega_z}\Tilde{r}_H \Tilde{z}(f\h_++g\h_\times)\cos(\omega T_\text{eff} ).
\end{equation}
Unlike the unperturbed case, where $g_{\mu\nu} k^\mu k^\nu$ at the acoustic horizon is equal to 0 (see Eq.\,\eqref{eq:k2unperturbed}), here we have a dynamical horizon and because of that, it is not required to be a null hypersurface \cite{dynamicalHor}.  Depending on the sign of the cosine we can have a null, timelike or spacelike hypersurface. However, for a Killing horizon this quantity should be zero and indeed for sufficiently low fluctuations this term vanishes: it is of order $\mathcal{O}(\epsilon \omega^2/\omega_r \omega_z)$. 
The null vector field computed in this appendix allows to evaluate the expansion rate, shear, and rotation of the null geodesic congruence\,\cite{2004rtmb.book.Poisson}. We will discuss this topic in a future publication.

\end{document}